\documentclass{article}

      \usepackage{fullpage}
\makeatletter

\newcommand{\Rmnum}[1]{\expandafter\@slowromancap\romannumeral #1@}
\makeatother

\usepackage{graphics} 
\usepackage[utf8]{inputenc} 
\usepackage[T1]{fontenc}    
\usepackage{hyperref}       
\usepackage{url}            
\usepackage{booktabs}       
\usepackage{amsfonts}       
\usepackage{nicefrac}       
\usepackage{microtype}      
\usepackage{lipsum}

\usepackage{graphicx}
\usepackage{epsfig} 
\usepackage{mathptmx} 
\usepackage{times} 
\usepackage{amsmath} 
\usepackage{amssymb}  
\usepackage{url}
\usepackage{cite}
\usepackage{color}
\usepackage{float}
\usepackage[english]{babel}
\usepackage{caption}
\usepackage{hhline}
\usepackage{multicol}
\usepackage{lipsum}
\usepackage{tikz}
\usepackage{epstopdf}
\usepackage{placeins}
\usepackage{dsfont}
\usepackage{mathtools}
\usepackage{bbm}
\usepackage{csquotes}
\usepackage{subcaption} 
\usepackage{authblk}

\newcommand*{\QEDA}{\hfill\ensuremath{\blacksquare}}%

\newtheorem{theorem}{Theorem}

\newtheorem{remark}{Remark}

\DeclareMathAlphabet{\mathcal}{OMS}{cmsy}{m}{n}
\SetMathAlphabet{\mathcal}{bold}{OMS}{cmsy}{b}{n}

\graphicspath{{./fig/}}

\title{\LARGE \bf Design of Robust Path-Following Control System for Self-driving Vehicles Using Extended High-Gain Observer \author{{Yasir K. Al-Nadawi, Hothaifa Al-Qassab, Daniel Kent, Su Pang, Vaibhav Srivastava, \\ and Hayder Radha}}
\thanks{This work has been supported in part by the General Motors and Society of Automotive Engineers (SAE) AutoDrive Challenge initiative, the Michigan State Foundation under the Strategic Partnership Program (SPG) and the Connected and Autonomous Networked-Vehicles for Active Safety (CANVAS) Research Program at Michigan State University.}
\thanks{Y. K. Al-Nadawi, H. Al-Qassab, D. Kent, S. Pang, V. Srivastava, and H. Radha are with the Department of Electrical and Computer Engineering, Michigan State University, East Lansing, MI 48824, USA. {\tt\small \{alnadawi, alqassab, kentdan3, pangsu, vaibhav, radha\}@egr.msu.edu}.}
}

\begin{document}

\date{}

\maketitle \thispagestyle{empty} \pagestyle{plain}

\begin{abstract}
In the real-world, self-driving vehicles are required to achieve steering maneuvers in both uncontrolled and uncertain environments while maintaining high levels of safety and passengers' comfort. Ignoring these requirements would inherently cause a significant degradation in the performance of the control system, and consequently, could lead to life-threatening scenarios. In this paper, we present a robust path following control of a self-driving vehicle under mismatched perturbations due to the effect of parametric uncertainties, vehicle side-slip angle, and road banking. In particular, the proposed control framework includes two parts. The first part ensures that the lateral and the yaw dynamics behave with nominal desired dynamics by canceling undesired dynamics. The second part is composed of two extended high-gain observers to estimate the system state variables and the perturbation terms. Our stability analysis of the closed-loop systems confirms exponential stability properties under the proposed control law. To validate the proposed control system, the controller is implemented experimentally on an autonomous vehicle research platform and tested in different road conditions that include flat, inclined, and banked roads. The experimental results show the effectiveness of the controller, they also illustrate the capability of the controller in achieving comparable performance under inclined and banked roads as compared to flat roads under a range of longitudinal velocities.        
\end{abstract}

\section{Introduction}\label{Sec:Intr}
Passengers and drivers' safety, traffic congestion, and increasing pollution levels around the globe have been the catalyst for the development of emerging intelligent transportation systems. In particular, autonomous driving is being targeted to improve safety, passengers’ comfort, fuel efficiency, and better utilization of transportation infrastructures \cite{7791799}. The ultimate goal is to achieve full autonomy in highly dynamic and uncertain urban and high-way environments within the next few years \cite{8400581}. Meanwhile, there are major technical challenges. Among these challenges is the design and realization of a robust path-following control system that can operate in a wide range of conditions. Current trajectory planning algorithms are designed using simplified vehicle models, which may violate certain constraints and/or fail to accommodate certain traffic scene conditions. Moreover, since these vehicles are required to serve in uncontrolled environments, they are quite vulnerable to a variety of uncertainty sources, such as difficult environmental conditions (e.g., high-slippage roads, ragged roads, low-visibility, etc.), measurement error, and modeling discrepancy \cite{6859253}.    

The steering control design problem for ground vehicles has been studied and reported in many papers. Various control methodologies such as linear model predictive control (MPC) \cite{6669331,8028942}, nonlinear robust model predictive control (NRMPC) \cite{6859253}, sliding mode control~\cite{7791799,327146,8751932} and backstepping control combined with sliding mode control \cite{8717647} have been developed for this problem. 
 In \cite{doi:10.1080/00423110701602696}, a steering controller based on differential flatness technique designed for lateral vehicle dynamics. Another flatness-based steering controller is designed in combination with an optimal State-Dependent Riccati Equation (SDRE) approach is presented in \cite{4162496}. Iterative learning control method was considered for the lateral vehicle dynamics model in \cite{7171151}. Robust steering control system is designed based on $\mathcal{H}_{\infty}$ approach \cite{8014072}.

Disturbance estimation based approaches are also considered in the literature. 
An active disturbance rejection control methodology is applied with the assumption that the vehicle is modeled with a lateral bicycle kinematic model in \cite{6640392}. In this work, the authors designed an extended state observer that estimates the system state variables and lumped disturbance terms, then a control law is designed utilizing the estimated states and disturbance terms to steer the vehicle to the desired path. This work has been extended in \cite{7410061} to incorporate the lateral bicycle dynamical model, wherein the authors obtained an equivalent model by establishing a flatness property for the linearized dynamics, then designed an adaptive ADRC based controller.   


In this paper, we consider the problem of lateral steering control of a steer-by-wire self-driving vehicle under the effect of vehicle side-slip, road banking, and the model parametric uncertainties. In order to design the controller, we assume that the vehicle is modeled with two degrees of freedom nonlinear bicycle dynamical model. This model by nature is underactuated due to having only one actuator input, which is the steering angle, and two degrees of freedom to control, which are the lateral position and the yaw angle. Moreover, the effect of vehicle side-slip, road banking in the dynamics in addition to parametric uncertainties create mismatched coupling disturbance terms that are function of the system state variables, input steering angle, and other perturbation terms. To design the controller, firstly, we derive the error dynamics with the consideration of Dugoff's nonlinear tire model. The next step is to design a control law based on the equilibrium balance concept \cite{Getz:1995:DIN:923055}. In order to estimate the unmeasured state variables and the disturbance terms, we designed two extended high-gain observers to recover the lateral position and the yaw angle dynamics, respectively. Finally, the estimated state variables and the disturbances terms are used in the control law to achieve the control objective.            

The work of \cite{6640392} and \cite{7410061} are comparable to our design due to the nature of the control law that utilizes a disturbance observer in the loop. However, our work can be differentiated in the following way
\begin{enumerate}
\item In \cite{6640392}, the assumed model is the kinematic bicycle model, which assumes a low longitudinal velocity. Moreover, the effect of vehicle side-slipping and the road angles were not considered. 
\item  In \cite{7410061}, the authors considered a linearized dynamical bicycle model, while in our case we are using a nonlinear model with the consideration of tire dynamics. Moreover, the road banking angle effect was not considered in their problem formulation. The other major difference, due to using the differential flatness theory, the state variables of the alternative equivalent model that is used to design the controller and the extended state observer are obtained in terms of the differential flat output and its time derivatives up to relative degree equal to 4, which accordingly requires the design of a fifth-order extended state observer to estimate higher-order derivatives. In our case, the controller allows us to design two separate extended third-order high-gain observers for the lateral position and the yaw angle subsystems. 
\item Our work conducts experimental implantation of the proposed controller, while in \cite{6640392} and \cite{7410061}, only a simulation study is considered. 
\end{enumerate}

The remaining of the paper is organized in the following way. In Section \ref{Sec:MPI}, the lateral dynamics of the vehicle is explained and the error dynamics with respect to the desired path are derived. The proposed controller and the extended high-gain observers' design are presented in Section \ref{Sec:Con_EHGO}. Stability analysis and main theoretical result are provided in Section \ref{Sec:Stability_Analysis}. Section \ref{Sec:Con_Imp} provides an overview of our research platform vehicle and illustrates the controller experimental implementation tests. Finally, the major conclusions of the work and future directions are stated in section \ref{Sec:Conclusion}. 
         
\section{Lateral Dynamics of The Vehicle Model} \label{Sec:MPI}
Under the assumption of symmetrical dynamical characteristics of the vehicle sides, the vehicle lateral dynamics can be expressed in body-fixed coordinates with a bicycle model with two degrees of freedom, which are the vehicles' lateral motion and the yaw angle dynamics. In this work, the longitudinal velocity is assumed to be controlled separately. By considering the force resulted due to the road banking angle, the lateral motion dynamics can be expressed as \cite{Raja2012}:
\begin{equation}\label{Eq:Lateral_mot_dyn}
m \ a_{y}=f_{yf}+f_{yr}+f_{b},
\end{equation}
where $m$ is the vehicle mass, $a_{y}$ is the vehicle inertial acceleration along the lateral axis $y$, and $f_{b}$ is the force induced due to the road banking, which can be calculated from $f_{b}=g\ \mbox{sin}\left(\phi\right)$, $\phi$ is the road banking angle and $g$ is the gravitational acceleration constant. $f_{yf}$ and $f_{yr}$ are the front and rear lateral tire forces, respectively, and are assumed generated by Dugoff's tire model, which is represented as\cite{Raja2012}:
\begin{equation}\label{Eq:Lateral_force_f_p}
    f_{yi}=\bar{f}_{yi}+\tilde{f}_{yi}, \quad i \in \{f,r\},
\end{equation}
where $\bar{f}_{yf}$ and $\bar{f}_{yr}$ are the nominal lateral forces terms for the front and rear tire, respectively. $\tilde{f}_{yf}$ and $\tilde{f}_{yr}$ are the corresponding perturbation terms, respectively, and can be expressed as:
\begin{equation}\label{Eq:fyf_tilde}
    \tilde{f}_{yi}=\tilde{C}_{i}  \frac{\mbox{tan}\left(\bar{\theta}_{li}\right)f\left(\gamma_i \right)}{1+\beta_{x}}-\bar{f}_{yi}, \quad i \in \{f,r\},
\end{equation}
where $\bar{\theta}_{lf}$ and $\bar{\theta}_{lr}$ are the front and rear tire side-slip angle, $\beta_{x}$ is the longitudinal slip ratio, $\tilde{C}_{f}$ and $\tilde{C}_{r}$ are the front and rear cornering stiffness coefficients, respectively. The function $\gamma$ is defined by
\begin{equation*}
    \gamma_i= \mu_{r} \frac{\left(1+\beta_{x}\right)f_{z}}{2\sqrt{\left(C_{x}\beta_{x}\right)^2+\left(\tilde{C}_{y}\mbox{tan}\left(\bar{\theta}_{li}\right)\right)^2}}, \quad i \in \{f,r\},\quad \text{and }
\end{equation*}

\begin{equation}
    f\left(\gamma\right)=\left\{\begin{array}{ccr}
        (2-\gamma) \gamma, & \ \ \ & \mbox{if}~ \gamma<1, \\
        1, & \ \ \  & \mbox{if}~ \gamma\geq 1,
    \end{array}\right.
\end{equation}
where $\mu_{r}$ is the tire-road friction coefficient, $\tilde{C}_{x}$ is the longitudinal tire stiffness coefficient, $f_{z}$ is the vertical force on the tire, and $\tilde{C}_{y}=\tilde{C}_{f}+\tilde{C}_{r}$.
The front and rear tire side slip angles can be expressed as~\cite{Raja2012}:
\begin{equation*}
\bar \theta_{l_{f}}=\delta-  \tan^{-1}\left(\frac{\dot{y}+l_{f}\dot{\psi}}{\nu_{x}}\right) \quad \text{and }
\bar \theta_{l_{r}}=- \tan^{-1} \left(\frac{\dot{y}-l_{r}\dot{\psi}}{\nu_{x}}\right),
\end{equation*} 
where $\delta$ is the steering angle, $\dot{y}$ is the velocity along the lateral axis $y$, $\psi$ is the yaw angle, and $\nu_{x}$ is the velocity along the longitudinal axis $x$. 

The front and rear lateral nominal tire forces can be approximated by linear functions of the corresponding tire slip angles utilizing the small slipping angle approximation as follows:
\begin{equation*}
\bar{f}_{yf}=2\ C_{f}\ \theta_{l_{f}} \quad \text{and} \quad \bar{f}_{yr}=2\ C_{r}\ \theta_{l_{r}},
\end{equation*}
where $C_{f}$ and $C_{r}$ are nominal front and rear tires cornering stiffness coefficients, respectively, and $\theta_{l_{f}}$ and $\theta_{l_{r}}$ are the approximate slip angles of the front and rear tires, respectively, and they can expressed as \cite{7154581}:
\begin{equation*}
\theta_{l_{f}}=\delta-\ \frac{\dot{y}+l_{f}\dot{\psi}}{\nu_{x}} \quad \text{and} \quad 
\theta_{l_{r}}=-\ \frac{\dot{y}-l_{r}\dot{\psi}}{\nu_{x}}. 
\end{equation*}  
Consider equation (\ref{Eq:Lateral_mot_dyn}), the inertial acceleration along the lateral axis $y$ is composed of two parts, the first one is the lateral acceleration $\ddot{y}$ and the second term is due to the centripetal acceleration \cite{Raja2012}:
\begin{equation}\label{Eq:Centri_accel}
a_{y}=\ddot{y}+\nu_{x}\ \dot{\psi}
\end{equation}
The second part of the model, which is the yaw angle dynamics is calculated by obtaining the momentum balance about the $z$-axis, yields
\vspace{-2 mm}
\begin{equation}\label{Eq:Yaw_dyn}
I_{z}\ \ddot{\psi}=l_{f}\ f_{yf}-l_{r}\ f_{yr},
\end{equation} 
where $I_{z}$ is the yaw moment of inertia of the vehicle, $l_{f}$ and $l_{r}$ are the distances of the front and rear tires from the vehicle center of gravity, respectively.
\begin{remark}
In the subsequent sections, we assume that the vehicle operates such that tire side-slip angles obeys the condition ($|\bar{\theta}_{li}|< \frac{\pi}{2}$).
It follows that in this domain $\tilde{f}_{yi}$ is continuously differentiable. We defer the rigorous analysis of the requirements on the desired trajectories and the control laws to ensure that this assumption holds to a future work. In our experimental validation, this assumption always holds. 
\end{remark}
\subsection{Error Dynamics in Terms of Desired Path}\label{Sec:Err_dyn}
For the purpose of controller design, it is useful to express the system dynamics with respect to the desired trajectory coordinates. Under the assumption of constant longitudinal velocity $\nu_{x}$ assumption \cite{327146}, the desired yaw rate is derived from the desired path in the following way \cite{Raja2012}:
\begin{equation}\label{Eq:Desired_Yaw_rate}
\dot{\psi}_{des}=\kappa_{s}\left(t\right)\ \nu_{x},
\end{equation}
where $\kappa_{s}\left(t\right)$ is the desired trajectory curvature function. Accordingly, the desired path lateral acceleration can be derived as \cite{327146}, \cite{Raja2012}:
\begin{equation}\label{Eq:desired_lat_acc}
a_{y_{des}}=\nu_{x}\ \dot{\psi}_{des}=\kappa_{s}\left(t\right)\ \nu_{x}^{2}.
\end{equation}
Let $e_{1}$ represents the distance from the center of gravity of the vehicle to the desired trajectory, then
\begin{equation}\label{Eq:ddot_e1}
\begin{split}
\ddot{e}_{1}&=a_{y}-a_{y_{des}}=\ddot{y}+\nu_{x}\ \dot{\psi}-\kappa_{s}\ \nu_{x}^{2} \\
            &=\ddot{y}+\nu_{x}\left(\dot{\psi}-\dot{\psi}_{des}\right).
\end{split}
\end{equation}
Define $e_{2}$ as the orientation error with respect to the desired trajectory, then
\begin{equation}\label{Eq:e2}
e_{2}=\psi-\psi_{des},
\end{equation}
and 
\begin{equation}\label{Eq:ddot_e1_2}
\ddot{e}_{1}=\ddot{y}+\nu_{x}\ \dot{e}_{2}.
\end{equation}
Let $z_{1}:=z_{1}$, $\dot{z}_{1}=\dot{e}_{1}:=z_{2}$, $z_{3}:=e_{2}$, and $\dot{z}_{3}=\dot{e}_{2}=z_{4}$. By using equations (\ref{Eq:e2}) and (\ref{Eq:ddot_e1_2}), the system dynamics can be represented with respect to the error coordinates as follows:
\begin{equation}\label{Eq:Err_dyn_1}
\begin{split}
\dot{z}_{1}&=z_{2} \\
\dot{z}_{2}&= a_{22}\ z_{2}+a_{23}\ z_{3}+a_{24}\ z_4+b_{21}\ \delta+ d_{l}\left(\cdot\right) \\
\dot{z}_{3}&=z_{4} \\
\dot{z}_{4}&=a_{42}\ z_{2}+a_{43}\ z_{3}+a_{44}\ z_4+b_{41}\ \delta+ d_{\psi}\left(\cdot\right),
\end{split}
\end{equation} 
where
\begin{equation*}
\begin{array}{cc}
a_{22}=-\left(\frac{2\ C_{f}+2\ C_{r}}{m\ \nu_{x}}\right), & a_{23}=-\nu_{x}\ a_{22}, \\
a_{24}=-\left(\frac{2\ C_{f}\ l_{f}-2\ C_{r}\ l_{r}}{m\ \nu_{x}}\right), & b_{21}=\frac{2\ C_{f}}{m}, \\
a_{42}=-\left(\frac{2\ l_{f}\ C_{f}-2\ l_{r}\ C_{r}}{I_{z}\ \nu_{x}}\right), & a_{43}=-\nu_{x}\ a_{42}, \\
a_{44}=-\left(\frac{2\ C_{f}\ l_{f}^{2}+2\ C_{r}\ l_{r}^{2}}{I_{z}\ \nu_{x}}\right), & b_{41}=\frac{2\ C_{f}\ l_{f}}{I_z},
\end{array}
\end{equation*}
and the disturbance inputs at the lateral and yaw dynamics are:
\begin{equation}\label{Eq:later_dist}
d_{l}\left(\cdot\right)=g\ \mbox{sin}\left(\phi\right)+\frac{\tilde{f}_{yf}+\tilde{f}_{yr}}{m}-\alpha\dot{\psi}_{des}
\end{equation}
\begin{equation}\label{Eq:Yaw_dist}
d_{\psi}\left(\cdot\right)=\frac{l_{f}\tilde{f}_{yf}-l_{r}\tilde{f}_{yr}}{I_{z}}-a_{44}\ \psi_{des}-\ddot{\psi}_{des},
\end{equation}
where $\alpha=\nu_{x}+a_{24}$.

\section{Controller Design Based on EHGO} \label{Sec:Con_EHGO}
Consider the error dynamics equation (\ref{Eq:Err_dyn_1}); by assuming that the system parameters are varying within specific known intervals. Therefore, the system parameters can be represented using the additive uncertainty form as follows:
\begin{equation}\label{Eq:Uncert_para}
\begin{split}
a_{ij} & = \bar{a}_{ij}+\Delta_{a_{ij}}, \ \ \ \ \ \, i=2,4 \ \, j=2,3,4, \\
b_{m1} & = \bar{b}_{m1}+\Delta_{b_{m1}}, \ \ \ \ \ \, m=2,4, 
\end{split}
\end{equation}
where $\bar{a}_{ij}$ and $\bar{b}_{m1}$ are the nominal parameters and $\Delta_{a_{ij}}$ and $\Delta_{b_{m1}}$ are the perturbation terms of the corresponding parameters, respectively. By using the forms (\ref{Eq:Uncert_para}), the error dynamics (\ref{Eq:Err_dyn_1}) can be rewritten as;
\begingroup\makeatletter\def\f@size{9}\check@mathfonts
\def\maketag@@@#1{\hbox{\m@th\normalsize\normalfont#1}}%
\begin{equation}\label{Eq:Err_dyn_2}
\begin{split}
\dot{z}_{1}&=z_{2} \\
\dot{z}_{2}&= \bar{a}_{22}\ z_{2}+\bar{a}_{23}\ z_{3}+\bar{a}_{24}\ z_4+\bar{b}_{21}\ \delta+ D_{l}\left(z,\delta,\Delta_{l},d_{l}\right) \\
\dot{z}_{3}&=z_{4} \\
\dot{z}_{4}&=\bar{a}_{42}\ z_{2}+\bar{a}_{43}\ z_{3}+\bar{a}_{44}\ z_4+\bar{b}_{41}\ \delta+ D_{\psi}\left(z,\delta,\Delta_{\psi},d_{\psi}\right),
\end{split}
\end{equation}
\endgroup
where $z=[z_{1}, z_{2}, z_{3}, z_{4}]^{T}$, $\Delta_{l}$ and $\Delta_{\psi}$ are terms resulted due to lumping parameters' perturbation terms, and $D_{l}$ and $D_{\psi}$ are the lateral and yaw overall perturbation terms, respectively.  
The error dynamics (\ref{Eq:Err_dyn_2}) suffer from being not only underactuated, but also highly coupled with uncertain terms function of state variables $z$, the steering input $\delta$, and other perturbation terms.
Therefore, the first step is to design a control law that makes the lateral subsystem ($z_{1}-z_{2}$) and the yaw angle subsystem ($z_{3}-z_{4}$) behave with nominal desired dynamics. In the subsequent steps, we assume that the state variables and the perturbation terms $D_{l}$ and $D_{\psi}$ are known and available from measurements. 

Consider the lateral subsystem of error dynamics (\ref{Eq:Err_dyn_2}), let the steering control input be designed as;
\begin{equation}\label{Eq:Control_delta}
\delta=\frac{1}{\bar{b}_{21}}\left[-\bar{a}_{22}\ z_{2}-\bar{a}_{23}\ z_{3}-\bar{a}_{24}\ z_{4}-D_{l}\left(\cdot\right)+u\right],
\end{equation} 
where $u$ is an auxiliary control input. Using the control law (\ref{Eq:Control_delta}), the error dynamics (\ref{Eq:Err_dyn_2}) is transformed into
\begingroup\makeatletter\def\f@size{9.5}\check@mathfonts
\def\maketag@@@#1{\hbox{\m@th\normalsize\normalfont#1}}%
\begin{equation}\label{Eq:Err_dyn_3}
\begin{split}
\dot{z}_{1}&=z_{2} \\
\dot{z}_{2}&= u \\
\dot{z}_{3}&=z_{4} \\
\dot{z}_{4}&=\alpha_{2}\ z_{2}+\alpha_{3}\ z_{3}+\alpha_{4}\ z_4+\frac{\bar{b}_{41}}{\bar{b}_{21}}u-\frac{\bar{b}_{41}}{\bar{b}_{21}}D_{l}\left(\cdot\right)+D_{\psi}\left(\cdot\right),
\end{split}
\end{equation} 
\endgroup
where $\alpha_{i}=\bar{a}_{4i}-\bar{a}_{2i}\frac{\bar{b}_{41}}{\bar{b}_{21}}, \ \ i=2,3,4$. Consider the above system (\ref{Eq:Err_dyn_3}), let 
\begingroup\makeatletter\def\f@size{8.8}\check@mathfonts
\def\maketag@@@#1{\hbox{\m@th\normalsize\normalfont#1}}%
\begin{equation}\label{Eq:Control_u}
u=\frac{\bar{b}_{21}}{\bar{b}_{41}}\left[-\alpha_{2}\ z_{2}-\alpha_{3}\ z_{3}-\alpha_{4}\ z_{4}+\frac{\bar{b}_{41}}{\bar{b}_{21}}\ D_{l}\left(\cdot\right)-D_{\psi}\left(\cdot\right)+u_{d}\right],
\end{equation}
where $u_{d}$ is another auxiliary control input. Utilizing (\ref{Eq:Control_u}) the system dynamics (\ref{Eq:Err_dyn_3}) will be converted to
\begin{equation}\label{Eq:Err_dyn_4}
\begin{split}
\dot{z}_{1}&=z_{2}, \ \ \ \ \dot{z}_{2}= u, \\
\dot{z}_{3}&=z_{4}, \ \ \ \ \dot{z}_{4}=u_{d}. \\
\end{split}
\end{equation} 
Consider the yaw subsystem ($z_{3}-z_{4}$) of system dynamics (\ref{Eq:Err_dyn_4}), let $u_{d}$ control law is designed as;
\begin{equation}\label{Eq:Control_ud}
u_{d}=-k_{3}\left(z_{3}-z_{3_{des}}\right)-k_{4}z_{4},
\end{equation}
where $k_{3}$ and $k_{4}$ are positive constants, $z_{3_{des}}$ is a desired reference that acts as a virtual control term to the $z_{1}-z_{2}$ subsystem. Similar to \cite{LEE2015146}, we follow the equilibrium balance concept introduced in \cite{Getz:1995:DIN:923055}. Define the set $\Sigma=\{z\in \mathcal{R}^{4}|z_{3}=z_{3_{des}},z_{4}=0\}$ and let the desired dynamics of the lateral ($z_{1}-z_{2}$) subsystem of (\ref{Eq:Err_dyn_4}) is 
\begin{equation}\label{Eq:desired_upper_dyn}
\begin{split}
\dot{z}_{1} & = z_{2} \\
\dot{z}_{2} & = \nu_{h} + \frac{\bar{b}_{21}}{\bar{b}_{41}}u_{d},
\end{split}
\end{equation}
where $\nu_{h}$ is a virtual control law, which will treated as vanishing perturbation term on (\ref{Eq:desired_upper_dyn}). In order to design the virtual controller $z_{3_{des}}$, we need to solve the following equality
\begin{equation}\label{Eq:Equal_cond}
\frac{\bar{b}_{21}}{\bar{b}_{41}}\left[-\alpha_{2}\ z_{2}-\alpha_{3}\ z_{3_{des}}+\alpha_{4}\ z_{4}+\frac{\bar{b}_{41}}{\bar{b}_{21}}D_{l}\left(\cdot\right)-D_{\psi}\left(\cdot\right)\right]=\nu_{h}.
\end{equation}
Then,
\begin{equation}\label{Eq:vc_z3des}
z_{3_{des}}=\frac{1}{\alpha}\left[-\alpha_{2}\ z_{2}-\alpha_{4}\ z_{4}+\frac{\bar{b}_{41}}{\bar{b}_{21}}D_{l}\left(\cdot\right)-D_{\psi}\left(\cdot\right)-\frac{\bar{b}_{41}}{\bar{b}_{21}}\nu_{h}\right].
\end{equation}
Generally, if $k_{3}$ and $k_{4}$ parameters are taken high enough, then ($z_{3}-z_{3_{des}}$) term and $z_{4}$ will converge quickly to zero, which will ensure condition (\ref{Eq:Equal_cond}) is satisfied. By choosing 
\begin{equation}\label{Eq:Nu_h_Control}
\nu_{h}=-\tau^{2} \eta_{1}\ z_{1}-\tau\eta_{2}\ z_{2},
\end{equation} 
for $\eta_{1},\eta_{2}>0$, and $0<\tau\leq 1$. The upper subsystem will be converted into
\begin{equation}
\begin{split}
\dot{z}_{1} & = z_{2} \\
\dot{z}_{2} & = -\tau^{2} \eta_{1}\ z_{1}-\tau\eta_{2}\ z_{2},
\end{split}
\end{equation}
In the previous steps, we assumed that all the state variables and the disturbance inputs are available for measurement, however, in practical scenarios, this is not the case. Therefore, an extended high-gain observer is designed to provide an estimation of both the state variables and the disturbance inputs under the assumption that both lateral position $y$ and the yaw angle $\psi$ are available for measurement. Consider the error dynamics (\ref{Eq:Err_dyn_2}), we design two extended high-gain observers for the upper and lower subsystems of the dynamics (\ref{Eq:Err_dyn_2}) based on the approach proposed in \cite{4668524} as follows;
\begin{equation}\label{Eq:EHGOs}
\begin{split}
\dot{\hat{z}}_{1} & = \hat{z}_{2}+\frac{h_{1}}{\epsilon}\left(z_{1}-\hat{z}_{1}\right) \\
\dot{\hat{z}}_{2} & = \bar{a}_{22}\ \hat{z}_{2}+\bar{a}_{23}\ \hat{z}_{3}+\bar{a}_{24}\ \hat{z}_{4}+\bar{b}_{21}\ \delta+\hat{D}_{l}+\frac{h_{2}}{\epsilon^{2}}\left(z_{1}-\hat{z}_{1}\right) \\
\dot{\hat{D}}_{l} & = \frac{h_{3}}{\epsilon^{3}}\left(z_{1}-\hat{z}_{1}\right) \\
\dot{\hat{z}}_{3} & = \hat{z}_{4}+\frac{g_{1}}{\epsilon}\left(z_{3}-\hat{z}_{3}\right) \\
\dot{\hat{z}}_{4} & = \bar{a}_{42}\ \hat{z}_{2}+\bar{a}_{43}\ \hat{z}_{3}+\bar{a}_{44}\ \hat{z}_{4}+\bar{b}_{41}\ \delta+\hat{D}_{\psi}+\frac{g_{2}}{\epsilon^{2}}\left(z_{3}-\hat{z}_{3}\right) \\
\dot{\hat{D}}_{\psi} & = \frac{g_{3}}{\epsilon^{3}}\left(z_{3}-\hat{z}_{3}\right), \\
\end{split}
\end{equation}
where $\hat{z}_{i}, \ \ \left(1\leq i\leq 4\right)$ are the estimated error variables, $\hat{D}_{l}$ and $\hat{D}_{\psi}$ are estimated lateral and yaw perturbation terms, the constants $h_{1}$, $h_{2}$, $h_{3}$, $g_{1}$, $g_{2}$, and $g_{3}$ are chosen such that the polynomials ($s^{3}+h_{1}s^{2}+h_{2}s+h_{3}$) and ($s^{3}+g_{1}s^{2}+g_{2}s+g_{3}$) are Hurwitz, $\epsilon$ is small positive constant. Accordingly, the control laws (\ref{Eq:Control_delta}), (\ref{Eq:Control_u}), (\ref{Eq:Control_ud}), (\ref{Eq:vc_z3des}), and (\ref{Eq:Nu_h_Control}) are implemented by replacing the error state variables and the disturbance input terms by their estimated surrogates. The steering angle output of control law (\ref{Eq:Control_delta}) is saturated in the implementation to prevent the observer peaking effect resulted due to the use of high-gains \cite{kha02}.  
\section{Stability of the Output Feedback Closed-Loop System}\label{Sec:Stability_Analysis}
Under the proposed output feedback control law (\ref{Eq:Control_u}), the main stability properties are summarized in the following theorem. 
\begin{theorem}\label{Th1}
Consider the closed-loop system dynamics formed of the error dynamics (\ref{Eq:Err_dyn_1}) under the control law (\ref{Eq:Control_delta}) and the high-gain observer dynamics (\ref{Eq:EHGOs}). Given that $\hat{z}\left(0\right)$ is bounded, then there exists $\tau^{\ast}>0$ and for every $\tau \in \left(o,\tau^{\ast}\right]$, there exists $\epsilon^{\ast}\left(\tau\right)>0$ such that for every $\epsilon \in \left(o,\epsilon^{\ast}\right]$ and for all initial conditions $z\left(0\right)$ in the interior of some compact set $\Omega$, the closed-loop system has an exponentially stable equilibrium point at $\left(z=0,~ \hat{z}=0\right)$.  
\end{theorem}
\textit{Sketch of the proof}: The proposed control methodology makes the $z_3$ error state converges to the virtual control $z_{3_{des}}$. Once ($z_{3}=z_{3_{des}}$), the ($z_{1}-z_{2}$) subsystem  will follow the desired dynamics (\ref{Eq:desired_upper_dyn}). The proof is done under assumption similar to the one considered in the proof of Theorem (6.7.6) of \cite{Getz:1995:DIN:923055}, in which the term $\nu_{h}$ in (\ref{Eq:desired_upper_dyn}) is treated as a small perturbation to the ($z_{1}-z_{2}$) subsystem. The proof is conducted in two parts, firstly, we show under the boundedness of $\hat{z}\left(0\right)$, that all the closed-loop trajectories are going to converge to a positively-invariant set that includes the origin and parameterized by the controller parameter $\tau$ and the high-gain observer $\epsilon$ in a finite time $t_{0}>0$ and stay therein forever. It should be noted that this set shrinks to zero as the parameters $\tau$ and $\epsilon$ tend to zero. In the second part, we construct a composite Lyapunov function candidate of the closed-loop variables and obtain its time-derivative along the solution of the closed-loop dynamics. Then, by establishing its negative definiteness, we can show that the closed-loop system has an exponentially stable equilibrium point at ($\left(z=0,~ \hat{z}=0\right)$).   $\QEDA$
\section{Controller Experimental Implementation and Validation}\label{Sec:Con_Imp}
\begin{figure}
	\centering
		\includegraphics[width=0.4\textwidth]{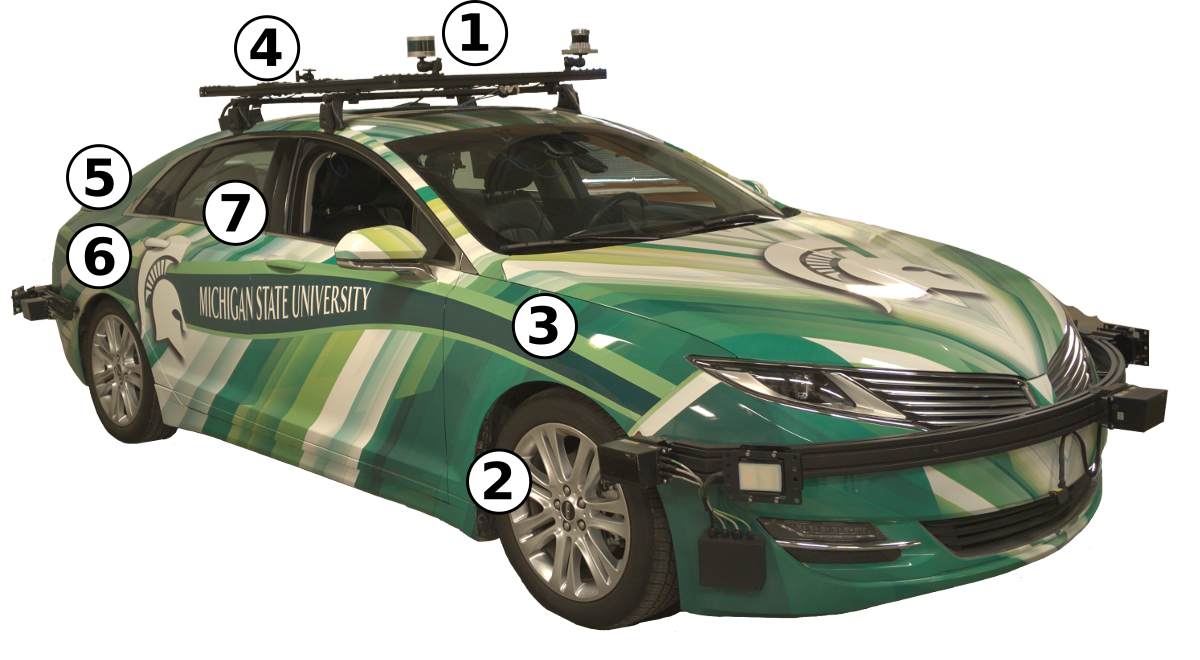}
		\caption{\small Autonomous vehicle research platform equipped with sensors and devices labeled as follows 1. Velodyne VLP-16 Lidar, 2. Wheel Encoders (inside wheel, stock), 3. Drive-By-Wire Kit (inside), 4. RTK GNSS Antenna (Mounting Location), 5. RTK GNSS Receiver (inside trunk), 6. Computation Platform (inside trunk), 7. Control Platform (inside passenger seat).}
		\label{fig:CanvasRP}
\end{figure} 
In this section, we provide an overview of the experimental implementation of the proposed controller. For this purpose, we use a drive-by-wire autonomous vehicle research platform equipped with sensors and devices described in Figure \ref{fig:CanvasRP}. The localization of the vehicle is performed using a technique based on pointcloud registration. The test platform vehicle is equipped with Lidar and GNSS navigation systems for localization. The navigation system uses the Universal Transverse Mercator (UTM) coordinate of WGS84 datum. Iterative Computed Point (ICP) registration method is used for pointcloud matching with pointcloud map. The 3D cloud maps used in our experiment were either generated by a professional surveying company or by ourselves with centimeter-level accuracy and are aligned with the UTM coordinate system. The lidar navigation system update rate is $10 \ \mbox{Hz}$ with centimeter-level accuracy.
The test platform performs localization in three steps. First, it takes an individual lidar scan and attempts to roughly transform the pointcloud data from the sensor’s coordinate frame into the frame of the loaded map. This process can use GNSS and/or prior scan matching results if they are available. Next, after the scan has been roughly aligned, the registration is refined using Normal Distributions Transform (NDT) registration, which is an extension of an earlier technique known as Iterative Computed Point (ICP) registration \cite{1249285}. Finally, once the mean-square error of the alignment falls below a certain threshold or the processing budget for alignment is otherwise exhausted, the resulting sensor to map transform is provided to the system, which can be used as an accurate estimate of the vehicle’s world-fixed position and rotation. This process repeats once new Lidar scan data is received. The entire localization process generally takes less than 12ms per scan under real-world conditions \cite{8690819}, which is less than the rate at which new scans are produced on our test platform (100ms). The controller module is coded with the Robotics System Toolbox under Simulink/Matlab package, which interfaces the controller with the Robotic Operating System (ROS-Kinetic) functionalities under Linux Ubuntu version 16.04 environment.  The laptop has Intel Core I-7 7700h and 16 GB of RAM. 
\begin{figure}
\centering
  \begin{subfigure}[pb!]{0.3\textwidth}
    \includegraphics[width=\textwidth,height=0.2\textheight]{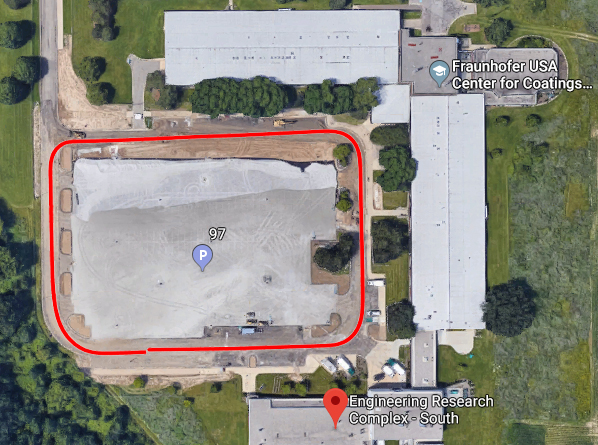}
    \caption{\small}
    \label{fig:ERC_1}
  \end{subfigure}
  \begin{subfigure}[bp!]{0.35\textwidth}
    \includegraphics[width=\textwidth]{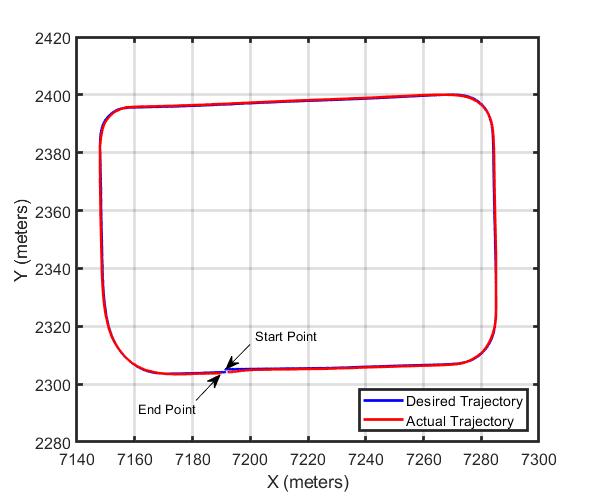}
    \caption{\small}
    \label{fig:ERC_2}
  \end{subfigure}
  \caption{\small Experimental results at test site 1 (a) Google maps screenshot of test site 1 overlaid with the desired trajectory. (b) Measured trajectory with respect to the desired path in inertial coordinates.}
\end{figure}\label{fig:ERC}
\begin{figure}[bp!]
	\centering
		\includegraphics[width=0.4\textwidth,height=0.22\textheight]{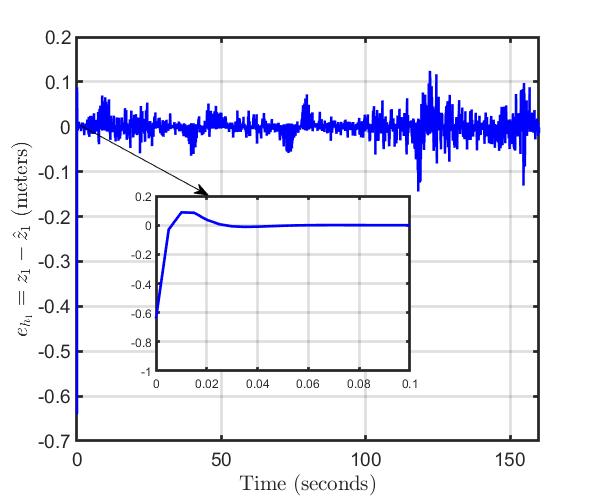}
		\caption{\small Estimation error of lateral position $e_{h_{1}}=z_{1}-\hat{z}_{1}$ for controller implementation on test site 1.}
		\label{fig:ERC_E_1}
\end{figure} 
\begin{figure}[bp!]
	\centering
		\includegraphics[width=0.4\textwidth,height=0.22\textheight]{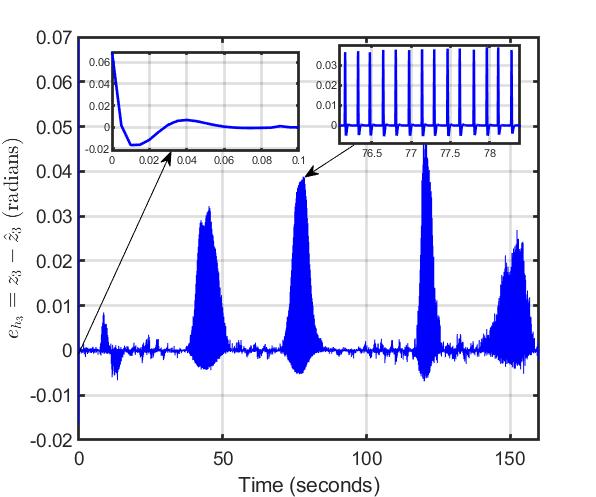}
		\caption{\small Estimation error of yaw angle $e_{h_{3}}=z_{3}-\hat{z}_{3}$ for controller implementation on test site 1.}
		\label{fig:ERC_E_2}
\end{figure}
\begin{figure}[bp!]
	\centering
		\includegraphics[width=0.4\textwidth,height=0.22\textheight]{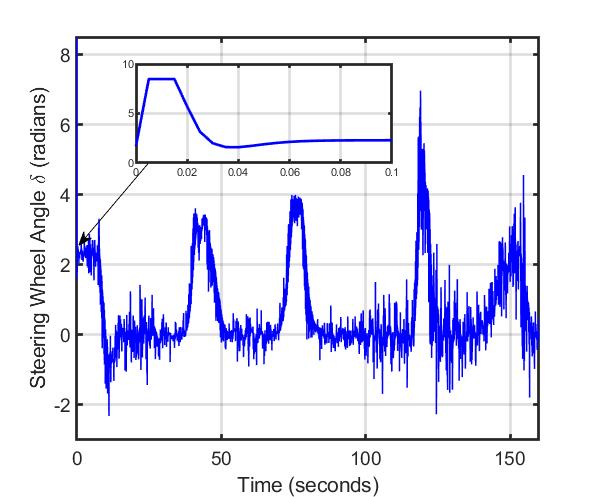}
		\caption{\small Steering angle profile for controller implantation on test site 1.}
		\label{fig:ERC_d}
\end{figure}
The experimental testing of the proposed controller took place in the following three sites; \textbf{Test Site 1}: Michigan State University campus (East Lansing, MI) - Engineering Research complex parking lot 97; \textbf{Test Site 2}: Michigan State University campus (East Lansing, MI) - Spartan Village (Middlevale road); and \textbf{Test Site 3}: Corrigan Oil (Spartan) Speedway (Mason, MI), where these test sites' google maps screenshots are shown in Figures \ref{fig:ERC_1}, \ref{fig:SPV_1}, and \ref{fig:SSW_1}, respectively. We utilize Site 1 which has a flat terrain to tune the controller and observer parameters and we use this set of parameters to test the controller on the other two sites. It is worth mentioning that in order to maintain high standards of safety requirements, we have implemented the testing with constant speed and the experiments run were aborted every time a pedestrian or other moving vehicles are in the vicinity of our testing vehicle. Testing site 2 and 3 are chosen due to having inclined and banked roads to challenge our closed-loop control system.  

The controller parameters are taken as; $\eta_{1}=2835000$, $\eta_{2}=31500$, $\tau=0.1$, $k_3=350000$, and $k_4=250000$. The high-gain observer parameters are set to; $h_1=g_1=2$, $h_2=g_2=1$, $h_3=g_3=0.5$, and $\epsilon=0.005$. The steering angle is saturated to ($\pm 2.7 \pi \ \mbox{radians}$) to buffer the control law from the observer peaking. In Figure \ref{fig:ERC_2}, the vehicle measured trajectory is plotted (in red) with respect to the desired trajectory (in blue) for test site 1 with vehicle longitudinal velocity is being set to $4.4704 \  \mbox{m/sec} \ \left(10 \ \mbox{mph} \right)$. The lateral and yaw angle estimation errors are shown in Figures \ref{fig:ERC_E_1} and \ref{fig:ERC_E_2}, respectively. Notice that both the estimation error converge in no more than $0.06 \ \mbox{sec}$ as shown in the magnified part of both figures at the initial transient response. However, a chattering appears clearly at certain intervals, especially for the yaw angle estimation error as shown in the magnified part at $t=77 \ \mbox{sec}$ of Figure \ref{fig:ERC_E_2}. This effect can be mitigated by tuning the controller parameters and/or decreasing the observer parameter $\epsilon$. However, due to numerical difficulty, we could not decrease the parameter $\epsilon$ below $0.005$. The steering angle is shown in Figure \ref{fig:ERC_d}, notice that the steering angle is saturated at the first ($0.002 \ \mbox{sec}$) as shown in the magnified part of the figure, this is due to the observer's peaking effect. 
\begin{figure}
\centering
  \begin{subfigure}[h!]{0.3\textwidth}
    \includegraphics[width=\textwidth,height=0.2\textheight]{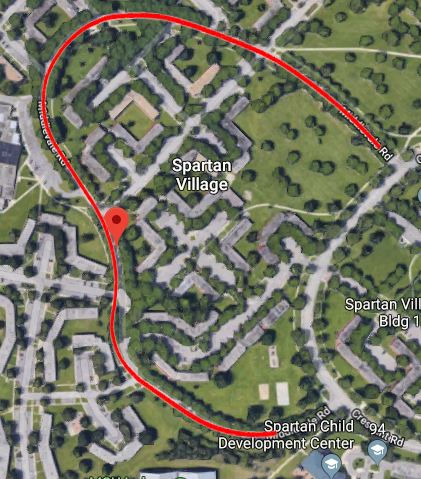}
    \caption{\small}
    \label{fig:SPV_1}
  \end{subfigure}
  \begin{subfigure}[h!]{0.35\textwidth}
    \includegraphics[width=\textwidth]{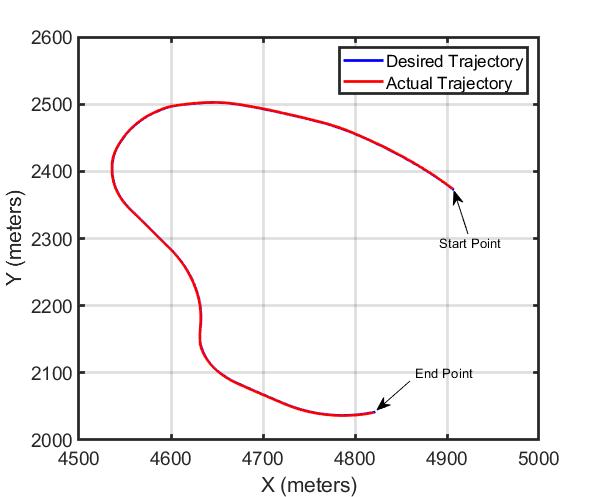}
    \caption{\small}
    \label{fig:SPV_2}
  \end{subfigure}
  \caption{\small Experimental results at test site 2 (a) Google maps screenshot of test site 2 overlaid with the desired trajectory. (b) Measured trajectory with respect to the desired path in inertial coordinates.}
\end{figure}\label{fig:SPV}
\begin{figure}[h!]
	\centering
		\includegraphics[width=0.4\textwidth,height=0.22\textheight]{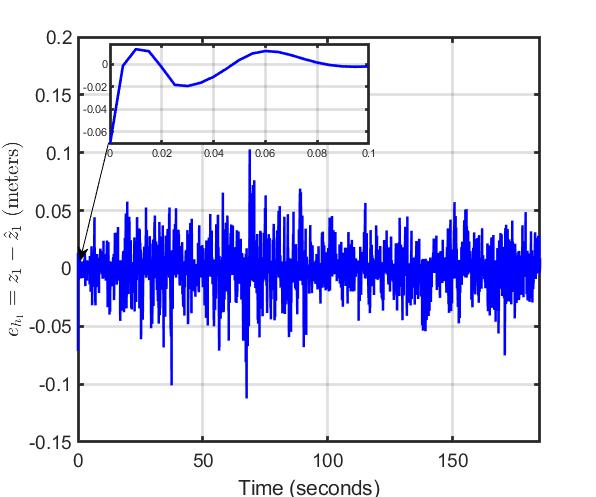}
		\caption{\small Estimation error of lateral position $e_{h_{1}}=z_{1}-\hat{z}_{1}$ for controller implementation on test site 2.}
		\label{fig:SPV_E_1}
\end{figure}
\begin{figure}[h!]
	\centering
		\includegraphics[width=0.4\textwidth,height=0.22\textheight]{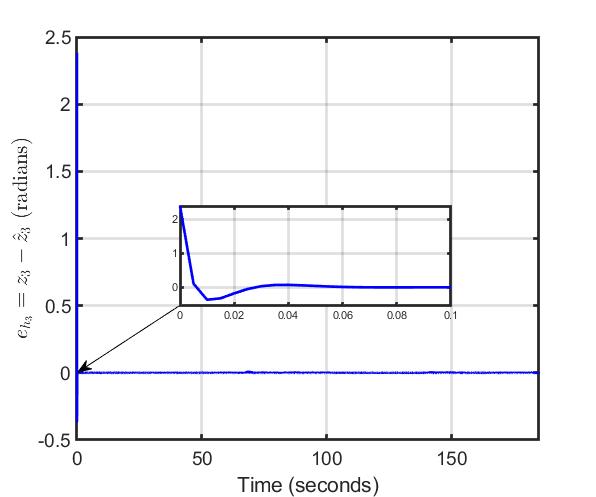}
		\caption{\small Estimation error of yaw angle $e_{h_{3}}=z_{3}-\hat{z}_{3}$ for controller implementation on test site 2.}
		\label{fig:SPV_E_2}
\end{figure}
\begin{figure}
	\centering
		\includegraphics[width=0.4\textwidth,height=0.22\textheight]{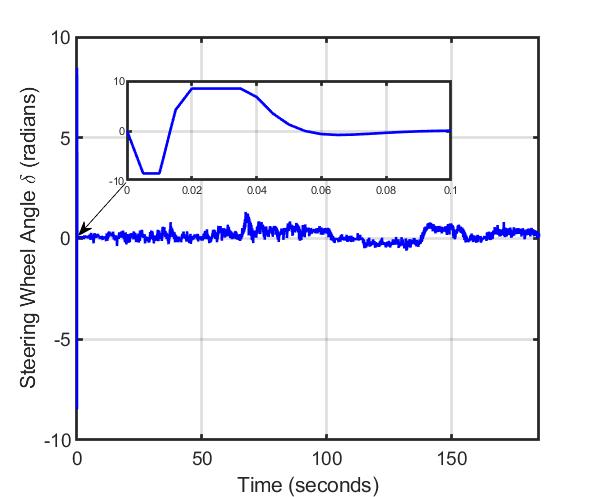}
		\caption{\small Steering angle profile for controller implantation on test site 2.}
		\label{fig:SPV_d}
\end{figure}

In the subsequent figures, the experimental results are demonstrated for both test sites 2 and 3, which have inclined and banked tracks compared to the flat terrain of test site 1. The longitudinal velocity of vehicle is chosen $6.7056 \  \mbox{m/sec} \ \left(15 \ \mbox{mph} \right)$ and $8.9408 \  \mbox{m/sec} \ \left(20 \ \mbox{mph} \right)$ for test sites 2 and 3, respectively. The choice of velocity values is done based on traffic scene limitations, speed limits, and safety considerations. It is noticed that using the same controller and observer parameters used for test site 1, the control system performance seems comparable to the results obtained on test site 1, although the geographical nature of the test tracks 2 and 3 are not flat due to being inclined and banked roads.       

By comparing the experimental results obtained from test site 2 and 3 to test site 1, it can be observed in Figures \ref{fig:SPV_E_1}, \ref{fig:SPV_E_2}, \ref{fig:SSW_E_1}, and \ref{fig:SSW_E_1} that the estimation errors also converge in around $0.06-0.08 \ \mbox{sec}$, which is a bit longer than what we observed with test site 1. Moreover, similar chattering phenomena is observed in observer responses as indicated in the magnified parts of Figures \ref{fig:SPV_E_2} and \ref{fig:SSW_E_2}. In Figures \ref{fig:SPV_d} and \ref{fig:SSW_d}, due to observer's peaking effect, the steering angle is saturated twice with the first $0.08 \ \mbox{secs}$ for the test site 2 case as shown in Figure \ref{fig:SPV_d}, while in the case of test 3, the saturation occurs only once as illustrated in figure \ref{fig:SSW_d}.  
\section{Conclusion}\label{Sec:Conclusion}
In this work, we designed and implemented an output feedback steering control law that takes into consideration the effect of multiple uncertain coupling terms in the dynamics. The control law (\ref{Eq:Control_delta}) is designed to stabilize the lateral position error dynamics subsystem and the the yaw angle error dynamics subsystem. This is done by a Backstepping-like approach, in which the yaw angle error variable are forced to track a virtual desired control input $z_{3_{des}}$. This virtual control by itself is designed to cancel nominal terms and the other perturbation terms. The unknown perturbation terms are estimated using two extended high-gain observers and their estimates are used within the control law to cancel out the effect of the perturbations in the lateral and yaw dynamics. Stability analysis shows that the proposed controller renders the output feedback closed-loop system exponentially stable. Experimental implementation demonstrates the efficacy of the proposed controller with inclined and banked test tracks.
\begin{figure}
\centering 
  \begin{subfigure}[h!]{0.3\textwidth}
    \includegraphics[width=\textwidth,height=0.2\textheight]{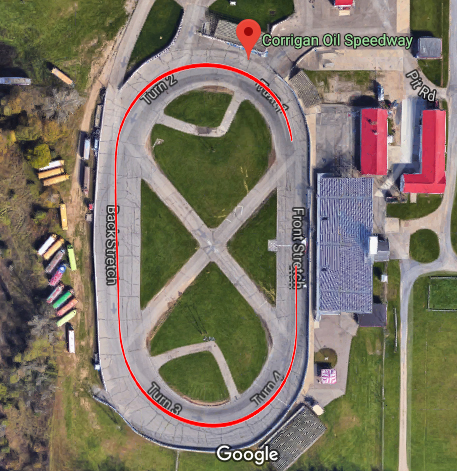}
    \caption{\small}
    \label{fig:SSW_1}
  \end{subfigure}
  \begin{subfigure}[h!]{0.35\textwidth}
    \includegraphics[width=\textwidth]{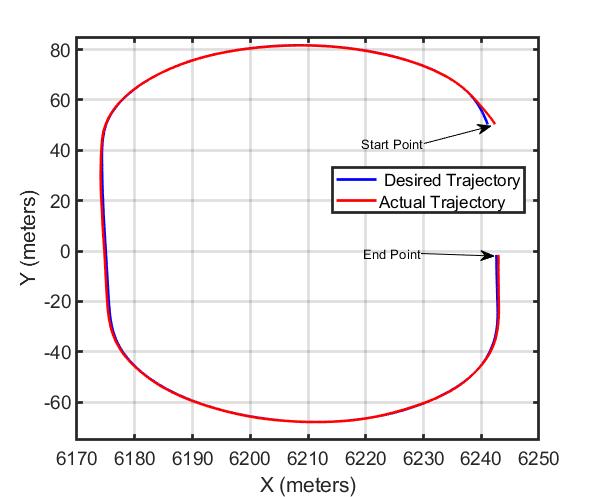}
    \caption{\small}
    \label{fig:SSW_2}
  \end{subfigure}
  \caption{\small Experimental results at test site 3 (a) Google maps screenshot of test site 3 overlaid with the desired trajectory. (b) Measured trajectory with respect to the desired path in inertial coordinates.}
\end{figure}\label{fig:SSW}
\begin{figure}[h!]
	\centering
		\includegraphics[width=0.4\textwidth,height=0.22\textheight]{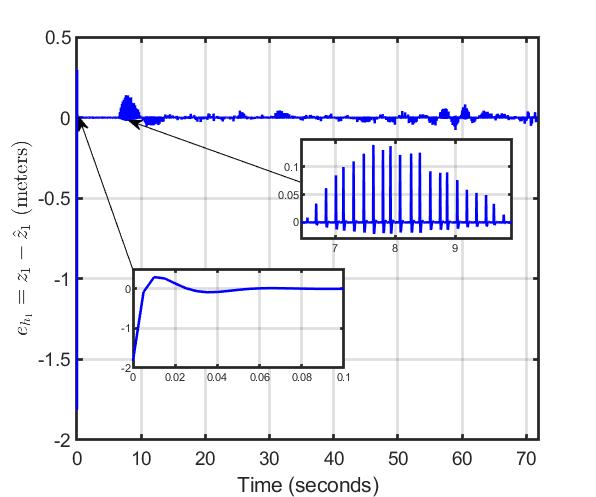}
		\caption{\small Estimation error of lateral position $e_{h_{1}}=z_{1}-\hat{z}_{1}$ for controller implementation on test site 3.}
		\label{fig:SSW_E_1}
\end{figure}
\begin{figure}[h!]
	\centering
		\includegraphics[width=0.4\textwidth,height=0.22\textheight]{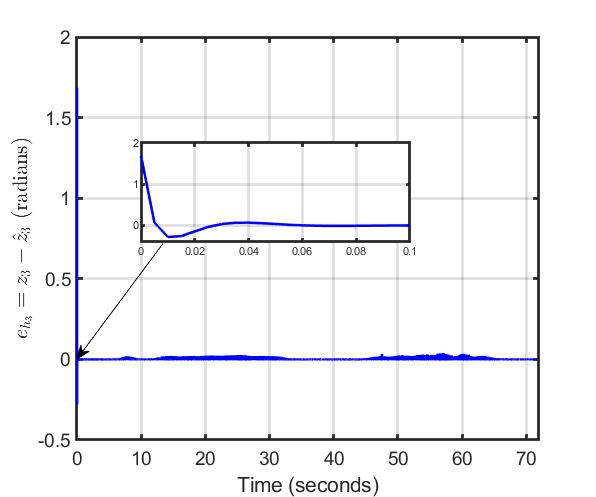}
		\caption{\small Estimation error of yaw angle $e_{h_{3}}=z_{3}-\hat{z}_{3}$ for controller implementation on test site 3.}
		\label{fig:SSW_E_2}
\end{figure}
\begin{figure}[h!]
	\centering
		\includegraphics[width=0.4\textwidth,height=0.22\textheight]{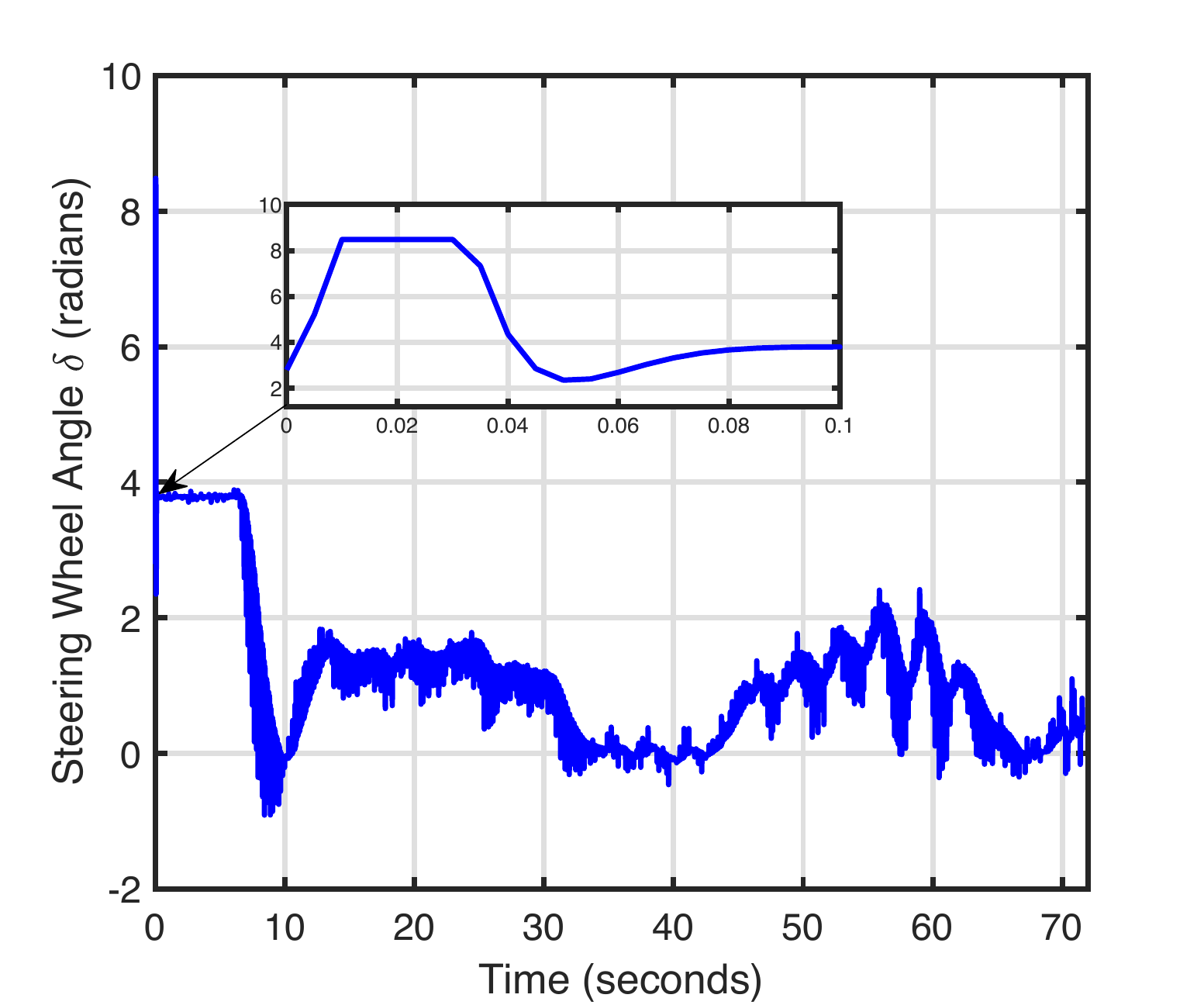}
		\caption{\small Steering angle profile for controller implantation on test site 3.}
		\label{fig:SSW_d}
\end{figure}
\section*{Acknowledgment}
We would like to thank the CARMERA team \cite{Carmera} who constructed the 3D pointcloud map of test site 2.

 \newcommand{\noop}[1]{}

\end{document}